\documentstyle{elsart}
                    \begin{document}
                    \begin{frontmatter} 
                    \title{The Landauer Resistance and Band Spectra For the
          Counting Quantum Turing Machine} 
                    \author{Paul Benioff}
                    \address{ Physics Division, Argonne National Laboratory
                     Argonne, IL 60439,
                     e-mail: pbenioff@anl.gov}

                    \begin{abstract}  
                    In other work, the generalized counting quantum Turing
          machine (GCQTM) was studied. For any $N$ this machine enumerates
          the first $2^{N}$ integers in succession as binary strings.  The
          generalization consists of associating a potential with read 1
          steps only.  The Landauer Resistance (LR) and band spectra were
          determined for the tight binding Hamiltonians associated with the
          GCQTM for energies below the potential height.  Here these
          calculations are extended to energies both above and below the
          barrier height.  For parameters and potentials in the electron
          region, The LR fluctuates rapidly between very high and very low
          values as a function of momentum. The rapidity and extent of the
          fluctuations increases rapidly with increasing $N$.  For $N=18$,
          the largest value considered, the LR shows good transmission
          probability as a function of momentum with numerous holes of very
          high LR values present. This is true for energies both above and
          below the potential height.  It is suggested that the main
          features of the LR can be explained by coherent superposition of
          the component waves reflected from or transmitted through or
          across the $2^{N-1}$ potentials present in the distribution.  If
          this explanation is correct, it provides a dramatic illustration
          of the effects of quantum nonlocality.
          \end{abstract}
          \end{frontmatter}

          \section{Introduction}
          Since the early 1980s quantum computation has been under active
          development in several directions.  This includes the development
          of efficient algorithms \cite{Shor,Grover,DeuJo}, simulation of
          physical systems\cite{Lloyd,Feynman1},  quantum error
          correction \cite{Shor1}, and use of quantum gates \cite{Lloyd1}.  
	  In  this work and in earlier work
          on Quantum Turing Machines \cite{Deutsch,Ben}, computations were
          assumed to proceed smoothly with no potentials present to cause
          reflections or decay of the transmitted state.

          The importance of including these potentials and the resulting
          degradation of performance has been emphasized by Landauer
          \cite{Landauer}. He has noted that potentials can be present
          because of environmental influences and physical limitations and
          errors \cite{Peres,Zurek} which can occur in realistic models.  

          In previous work \cite{Benioff1,Benioff2} on quantum Turing
          machines (QTMs) potentials were introduced by associating them
          with different types of steps.  Isolation from the environment is
          assumed.  In this case the Feynman \cite{Feynman} Hamiltonian for
          these generalized quantum Turing machines (GQTM)s,  can be also
          be described as a sum of tight binding Hamiltonians each of which
          has a different potential distribution along a one dimensional
          path of computation states.  Which of these Hamiltonians applies
          to the GQTM is determined by the initial state.

          The effect of these potentials was studied in detail for the
          counting generalized QTM (GQTM), which enumerates the nonnegative
          integers in succession as binary strings.  The potential
          distribution associated with this GQTM,  is closely related to
          the heirarchical sequence \cite{Benioff1,HuKe,IgTu}.  This
          sequence gives the number of $1s$ appearing before the first $0$
          in the successive iteration of integers represented as binary
          strings. 

          Both the band spectra and the Landauer Resistance
          \cite{Landauer1,Erdos} were calculated as a function of energy
          for computations which enumerate the first $2^{N}$ integers as
          binary strings. The calculations were limited to energies below
          the height of the potential (bound state energies).  

          In this paper the calculations made for the band spectra and LR
          for the counting GQTM will be extended by removal of the
          limitation of the energy range to energies below the barrier.
          Since the physical model used corresponds to head motion on a 1-D
          lattice of qubits with a Hamiltonian constructed from step
          operators, the full energy range remains finite.  



          In the next section the physical model and main points of the
          description of the Hamiltonians for GQTMs will be reviewed. The
          use of the transfer matrix to calculate both the band spectra and
          LR will be briefly reviewed in Section \ref{TM}.  Following this
          the counting GQTM will be briefly described.  Since the material
          has already been presented elsewhere, the discussion will be
          quite brief with additional material incorporated by reference.

          Results of calculations of the band spectra and the LR will be
          presented and discussed in Section \ref{BSLR}.  The LR, plotted
          as the $\log LR$ as a function of momentum, will be seen to
          fluctuate rapidly between very large and very small values.  The
          fluctuations get more rapid and extreme as $N$ is increased and
          the LR increases as the potential height is increased.  The
          association between the LR and band spectra for the associated
          periodic system obtained here is similar to that obtained by
          others \cite{Roy1,Roy2} in that peaks are associated with gaps 

          In Section \ref{disc} problems with extending the calculations to
          moderate values of $N$ are briefly discussed.  The possible
          importance of coherent superposition of wave components reflected
          from the many potentials in the distribution in producing the
          observed fluctuations in the LR will be discussed.  It is noted
          that this provides a good example of quantum nonlocality.

          \section{Hamiltonians for GQTMs}
          \label{HGQTM}

          \subsection{The Physical Model}
          The physical model for QTMs consists of a two-way infinite
          lattice of systems each of which can assume any state in a finite
          dimensional Hilbert space ${\cal H}_{s}$.  If ${\cal H}_{s}$ is
          two dimensional the systems are called qubits,  However this term
          is used here irrespective of the dimensionality of ${\cal
          H}_{s}$.   A head, described by a finite basis of states $\vert
          l\rangle$ with $l\epsilon L$, moves along the lattice interacting
          with the lattice systems at or adjacent to the head position.
          Possible elementary actions include any combination of head
          motion one site in either direction, lattice system state change,
          and head state change.

          The system states are all assumed to lie in a separable Hilbert
          space $\cal H$.  A particular basis, the computation basis,
          defined by the set of states $\{\vert l,j,S\rangle\}$ and which
          spans $\cal H$, is used.  Here $l,j$ refer to the internal state
          and lattice position of the head.  The qubit lattice computation
          basis state $\vert S\rangle =\otimes_{m=-\infty}^{\infty} \vert
          S(m)\rangle$ where $S$ is a function from the integers to the
          qubit state labels (e.g. $\{0,1\}$) such that  $S(m)\neq 0$ for
          at most a finite number of values of $m$.  This condition, the
          $0$ tail state condition, is one of many that can be imposed to
          keep the basis denumerable.  

          \subsection{Tight Binding Hamiltonians for GQTMs}
          A bounded linear operator $T$ is associated with each GQTM.  $T$
          or its adjoint is referred to as a step operator for the GQTM
          because iteration of $T$ (or its adjoint) corresponds to the
          successive steps in the forward (or backward) time direction of
          the GQTM. An infinitesimal time interval is associated with the
          steps of $T$ so that it can be used directly to construct a time
          independent Hamiltonian according to Feynman's prescription
          \cite{Feynman}:
          \begin{equation}
          H=K(2-T-T^{\dag}) \label{ham}
          \end{equation}
          where $K$ is a constant.  Note that if $T=Y$ where $Y$ is the
          bilateral shift along the lattice, then $H$ is the kinetic energy
          of free head motion on the lattice.  As such it is equivalent to
          the symmetrized discrete version of the second derivative,
          $(-\hbar^{2}/2m)d^{2}/dx^{2}$. 

          For each GQTM the associated step operator $T$ is defined as a
          finite sum over elementary step operators.  That is
          $T=\sum_{l,s}T_{ls}$ where the sum is over all head state labels
          $l\epsilon L$ and all qubit state labels $s\epsilon \cal S$ where
          $\cal S$ is the set of qubit state labels.  For binary qubits
          $\cal{S} =\{0,1\}$.  $T_{l,s}$ corresponds to the action taken by
          the GQTM associated with $T$ when the head in state $\vert
          l\rangle$ sees or reads a qubit in state $\vert s\rangle$.

          For each $l,s$ $T_{l,s}=\gamma_{l,s}W_{l,s}$ where $\gamma_{l,s}$
          is a positive real constant $\leq 1$.  $W_{l,s}$ is a sum over
          all lattice positions of products of projection operators, bit
          and head state change, and head position change operators.  The
          specific definition of $W_{l,s}$ appears elsewhere
          \cite{Benioff1}.

          The generalization from QTMs to GQTMs consists in allowing values
          of $\gamma_{l,s} \neq 1$.  Thus every QTM is a GQTM with
          $\gamma_{l,s}=1$ for all $l,s$. Note that the generalization does
          not expand or change the definition of quantum computability.

          The main condition imposed on $T$ and its adjoint is that they be
          distinct path generating in some basis $B$ which spans $\cal H$. 
          This means that iterations of $T$ or $T^{\dag}$ on any basis
          state in $B$, generate a path of states that, up to
          normalization, are also states in $B$.  The paths are distinct in
          that no two paths join, intersect, or branch.  

          Although $B$ can be any basis which spans $\cal H$, (see
          \cite{Benioff} for details), unless explicitly specified
          otherwise, $B$ is limited here to be the computation basis, 
          $\{\vert l,j,S\rangle \}$.  A path in the $B$ is defined as a set
          of states in $B$ ordered by iteration of $T$ or $T^{\dag}$.  If
          the state $\vert l_{m},j_{m},S_{m}\rangle =\vert m,i\rangle$ is
          the $mth$ state in some path $i$, then the $m+1st$ and $m-1th$
          states in the path are given respectively by $\vert m+1,i\rangle
          = W_{l_{m},S_{m}(j_{m})}\vert l_{m},j_{m},S_{m}\rangle = \vert
          l_{m+1},j_{m+1},S_{m+1}\rangle$ and $\vert m-1,i\rangle =
          W^{\dag}_{l_{m-1},S_{m-1}(j_{m-1})}\vert l_{m},j_{m},S_{m}\rangle
          = \vert l_{m-1},j_{m-1},S_{m-1}\rangle$.   These equations must
          be modified if either $T$ or $T^{\dag}$ annihilate $\vert
          m,i\rangle$.

          If $T$ is distinct path generating in $B$ then $T$ can be
          decomposed into a direct sum $T=\sum_{i}T_{i}P_{i}$ where $P_{i}$
          projects out a subspace ${\cal H}_{i}$ of $\cal H$ spanned by a
          subset $B_{i}$ of $B$. For each $i$ iteration of $T$ and
          $T^{\dag}$ on the states in $B_{i}$ orders them into a path,
          denote the computation path $i$. The action of $T_{i}$ and
          $T^{\dag}_{i}$ on each state $\vert m,i\rangle =\vert
          l_{m},j_{m},S_{m}(j_{m})\rangle$ which is not a terminal path
          state is given by 
          \begin{eqnarray}
          T\vert l_{m},j_{m},S_{m}\rangle & = &
          \gamma_{l_{m},S_{m}(j_{m})}\vert l_{m+1},j_{m+1}, S_{m+1} \rangle
          \nonumber \\
          T^{\dag}\vert l_{m},j_{m},S_{m}\rangle & = & \gamma_{l_{m-
          1},j_{m-1},S_{m-1}(j_{m-1})}\vert l_{m-1},j_{m-1},S_{m-1}\rangle.
          \label{TTdagcomp}
          \end{eqnarray}
          Additional details on the decomposition and its description as a
          sum of weighted shifts is given elsewhere \cite{Benioff2}.

          Corresponding to the decomposition of $T$ is a decomposition of
          the Hamiltonian, Eq. \ref{ham},  $H=\sum_{i}H_{i}P_{i}$ where for
          each $i$, 
          \begin{equation}
           H_{i}=K(2-T_{i}-T^{\dag}_{i}). \label{tbhi}
          \end{equation}
          As is noted elsewhere \cite{Benioff1} $H_{i}$ can also be written
          in the form
          \begin{equation}
          H_{i} = K(2-U_{i}-U^{\dag}_{i}) +V_{i} \label{tbhiv}
          \end{equation}
          where $U_{i}$ is a shift along the path $i$.  $U_{i}$ and
          $U^{\dag}_{i}$ are defined by Eqs. \ref{TTdagcomp} with $\gamma
          _{l,s} = 1$ for all $l,s$.  

          $K(2-U_{i}-U^{\dag}_{i})$ is, formally,  the kinetic energy
          associated with the evolution of the computation along path $i$. 
          For the physical model used here it is also the kinetic energy of
          head motion on the qubit lattice. (The kinetic energy depends on
          the magnitude but not the direction of the head momentum on the
          lattice.) 

          The potential $V_{i}=K(U_{i}-T_{i}+U^{\dag}_{i}-T^{\dag}_{i})$ is
          a nearest neighbor off-diagonal potential with matrix elements on
          internal path states given by 
          \begin{equation}
          \langle m^{\prime},i\vert V_{i}\vert m,i\rangle = K[(1-
          \gamma_{l_{m},S_{m}(j_{m})})\delta_{m^{\prime},m+1}+(1-
          \gamma_{l_{m-1},S_{m-1}(j_{m-1})}\delta_{m^{\prime},m-1}].
          \label{vjj}
          \end{equation}

          In Schr\"{o}dinger evolution under the action of $H$, Eq.
          \ref{ham}, the choice of which $H_{i}$ is active is determined by
          the initial state $\Psi(0)$.  If $\Psi(0)$ is a superposition of
          of computation basis states in one path $i$, then $H_{i}$ is the
          only  active component.  If $\Psi(0)$ is a superposition of
          computation basis states in different paths, then all $H_{i}$ are
          active for just those paths $i$ in which $\Psi(0)$ has a nonzero
          component.   In this case the states in the different paths
          evolve coherently which corresponds to quantum parallel
          computation \cite{Deutsch}.

          \section{Transfer Matrix}
          \label{TM}
          For each computation path $i$ one is interested in the evolution
          of the system state along the path under the action of $H_{i}$. 
          For paths in which the region of nonzero potentials is finite in
          extent with potential free regions on both sides,  the
          probability of completion of the computation is given by the
          transmission probability of the state through the potential
          region. This can be obtained by use of the transfer matrix
          following methods developed by Erdos and Herndon \cite{Erdos}.

          To be specific, assume the potential is nonzero in a region
          extending from state $\vert a,i\rangle$ to $\vert a+m,i\rangle$
          and is zero elsewhere on an infinite path $i$.  The states on
          either side of the region can be written in the form,
          \begin{eqnarray}
          \Psi_{I} & = & \sum_{j=-\infty}^{a-
          1}(A\mathrm{e}^{\mathrm{i}kj}+B\mathrm{e}^{-\mathrm{i}kj})\vert
          j\rangle \nonumber \\
          \Psi_{III} & = &
          \sum_{j=a+m+1}^{\infty}(F\mathrm{e}^{\mathrm{i}kj}+G\mathrm{e}^{-
          \mathrm{i}kj})\vert j\rangle \label{PsiIMF}
          \end{eqnarray}
          Here $j$ is the path state label (the path label $i$ in $\vert
          j,i\rangle$ is suppressed) and $k$ is the momentum in the
          potential free regions.  

          The form of $\Psi_{II}$, the state in the region II of nonzero
          potential, depends on the form of the potential. In general
          $\Psi_{II}=\sum_{j=a}^{a+m}(C_{j}\mathrm{e}^{\mathrm{i}l_{j}j}+D_
          {j}\mathrm{e}^{-\mathrm{i}l_{j}j})\vert j\rangle$ where the
          momenta $l_{j}$ and coefficients $C_{j},\; D_{j}$ depend on the
          potential height, Eq, \ref{vjj}, at $\vert j\rangle$.  Both the
          momenta and coefficients are constant over subregions of constant
          potential height. For any subregion for which the energy
          corresponding to the momentum $l_{j}$ is less than the potential
          height for the subregion, $l_{j}$ is replaced by
          $\mathrm{i}l_{j}$. Additional details are given elsewhere
          \cite{Erdos,Benioff2}.

          The four complex coefficients $A,B,F,G$ are completely determined
          by boundary conditions for the case under study and the
          properties of the transfer matrix.  For computations it is
          convenient to let Regions I and III be the initial and final
          state regions. In this case, the $A,\; B,\; F$ 
          coefficient terms of Eq. \ref{PsiIMF} are respectively the
          initial incoming state, the reflected term and the transmitted
          term. In this case $G=0$ as there is no incoming term  in the
          final state region.  Different boudary conditions are used if
          $\Psi$ represents an eigenstate of $H_{i}$.

          The transfer matrix $Z$ relates the coefficients $F,G$ to $A,B$
          according to \cite{Erdos}
          \begin{equation}
           \left| \begin{array}{c}
          F \\ G \end{array} \right). =Z  \left|
          \begin{array}{c}
          A \\ B \end{array} \right).  \label{Zdef}
          \end{equation}
          $Z$ is a unimodular (determinant=1) $2\times 2$ matrix which also
          satisfies \cite{Erdos}
          \begin{eqnarray}
          Z_{11} & = & Z^{*}_{22} \nonumber \\
          Z_{12} & = & Z^{*}_{21}. \label{Zprop}
          \end{eqnarray} 

          The above definitions can be iterated to define $Z$ as a product
          of transfer matrices for individual subregions of constant
          potential.  If there are $M$ such regions then $Z$ is the product
          of $M$ $2\times 2$ matrices, taken in the order in which the
          subregions occur in region II \cite{Erdos}.

          The Landauer Resistance (LR) \cite{Landauer1,Erdos}, which
          measures the resistance to transmission through the potential
          region, is defined by $LR=|B|^{2}/|F|^{2}$ with $G=0$.  For a
          computation on a path $i$, the probability of completion is given
          by $1/(1+LR)=|F|^{2}$. The properties of $Z$ can be used to
          obtain \cite{Erdos}
          \begin{equation}
          LR= |Z|_{12}|^{2} 
          \end{equation}
          It is clear from the above that the LR depends both on the path
          and on the system energy (i.e. the energy of the head as it moves
          on the lattice).

          The transfer matrix is also used to determine for each path $i$
          the energy band spectra for the infinite 1-D crystal whose unit
          cell is given by the potential distribution of $H_{i}$, Eq.
          \ref{vjj} (the associated periodic system).  As is well known
          \cite{Others,Roy1,Roy2} the band and gap distribution is obtained
          from a plot of $|TrZ|/2$ as a function of momentum or energy. 
          Regions in which $|TrZ|/2\leq 1$ correspond to bands; regions
          with $|TrZ|/2 >1$ correspond to gaps.

          \section{The Counting GQTM}
          \label{CGQTM}
          An interesting example which can be used to illustrate the
          foregoing is the counting QTM \cite{Benioff1,Benioff2}.  Starting
          with a string of $0s$ this machine generates nonnegative integers
          as binary strings in succession by repetitve addition of $1 \bmod
          2$.  One marker locates the start of the strings on the qubit
          lattice and another marker, if present, locates the maximum
          length of the strings.

          A step operator $T$ which carries this out can be given as a sum
          of 7 terms. A simple but nontrivial representation as a
          generalized QTM is obtained by setting, for all $l$,
          $\gamma_{l,s} = 1$ for all $s \neq 1$ and $\gamma_{l,1} =\gamma
          <1$.   The resulting step operator is given by 
          \begin{eqnarray}
          T & = & \sum_{j=-\infty}^{\infty}(Q_{0}P_{0j}uP_{j}
          +wQ_{0}P_{2j}uP_{j}+Q_{1}P_{0j}uP_{j}   \nonumber \\
           & & \mbox{}+wQ_{1}P_{2j}u^{\dag}P_{j} +\gamma
          Q_{2}v_{xj}P_{1j}u^{\dag}P_{j}+w^{\dag}Q_{2}v_{xj}P_{0j}uP_{j}  
          \nonumber \\
           & & \mbox{}+wQ_{2}P_{2j}uP_{j}) \label{Tex}
          \end{eqnarray}
          The projection operators $Q_{l},\,P_{s,j}\,P_{j}$ refer to the
          head in state $\vert l\rangle$, the site $j$ qubit in state
          $\vert s\rangle$, and the head at site $j$.  $w$ is a shift mod 3
          on the three head states ($wQ_{m}=Q_{m+1}w \bmod 3$) and $u$
          shifts the head along the lattice by one site
          ($uP_{j}=P_{j+1}u$).  The need for markers is accounted for here
          by choosing the qubits in the lattice to be ternary with states
          $\vert 0\rangle ,\vert 1\rangle ,\vert 2\rangle$. $\vert
          2\rangle$ is used as a marker and $\vert 0\rangle ,\vert
          1\rangle$ are used for binary strings.  The qubit transformation
          operator $v_{xj}=\sigma_{xj}(P_{0j}+P_{1j})+P_{2j}$ exchanges the
          states $\vert 0\rangle , \vert 1\rangle$ and does nothing to the
          state $\vert 2\rangle$ for the site $j$ qubit. 

          The adjoint $T^{\dag}$ is defined from $T$ in the usual way
          noting that the operators for the head states, head position
          states, and qubit states commute with one another.  Note that
          $v_{xj}^{\dag}=v_{xj}$.  The explicit form of $T^{\dag}$ as well
          as other aspects of this GQTM are given elsewhere
          \cite{Benioff2}.

          For this GQTM motion from state $\vert m,i\rangle$ to state
          $\vert m^{\prime},i\rangle$ is potential free unless the $5th$
          terms of $T$ or $T^{\dag}$ are active.  In this case, for $\vert
          m,i\rangle =\vert l_{m},j_{m}, S_{m}(j_{m})\rangle$ $l_{m}=2$ and
          $S_{m}(j_{m}) =1$ or(and) $l_{m-1}=2$ and $S_{m-1}(j_{m-1}) =1$. 
          If either one but not both of these conditions hold, then by Eq,
          \ref{vjj}, $V=K(1-\gamma)$.  If both conditions hold then
          $V=2K(1-\gamma)$.  Equivalent conditions can be given for the
          state $\vert m^{\prime},i\rangle$.  

          An analysis of the types of potential barriers that can occur in
          a distribution for this shows that they are at least 2 path
          states wide. A barrier that is $m+1$ sites wide, corresponding to
          $m$ successive (read 1) steps in which the $5th$ term of $T$ or
          $T^{\dag}$ is active, has a central core of height $V=2K(1-
          \gamma)$ occupying $m-1$ successive path sites or states.  The
          core is flanked by two potentials of height $V=K(1-\gamma)$
          occupying just one site or state.  For $m=1$ the barrier occupies
          2 sites and is of height $K(1-\gamma)$. 

          For potential free regions the relation between energy and
          momentum is given by $E=2K(1-\cos k)$.  For the core region of a
          barrier $E=2k(1-\gamma \cos{l})$ (or $E=2K(1-\gamma \cosh{l})$ if
          $E<V$). From these equations one obtains $\cos{k} =\gamma \cos{l}
          (\mbox{or} \cos{k}=\gamma \cosh{l})$.  For the flank regions
          $E=K(2-(\gamma ^{2}+2\gamma\cos{2h}+1)^{1/2}$ but this relation
          turns out not to be needed.

          The above describes all possible types of barriers that can occur
          on any computation path. However the distribution of barriers
          according to their width and spacing between barriers can vary
          widely among different paths.

          To apply this to a specific example, consider the initial state,
          shown in Figure \ref{fig:fig1}, with the head in state $\vert
          0\rangle$ in a wave packet localized to the left of the origin. 
          All qubits are in state $\vert 0\rangle$ except those at sites
          $0,N+1$ which are in state $\vert 2\rangle$.  The initial head
          and lattice qubit state are the path labels for this example. 
          Iteration of $T$ on this state generates in turn all the integers
          as binary strings of length $\leq N$.  When the space between the
          two markers is completely filled with $1s$, corresponding to the
          integer $2^{N}-1$, the last pass of the head changes all $1s$ to
          $0s$. The head in state $\vert 1\rangle $ then moves to the right
          away from the marker region as the enumeration is completed.  A
          more detailed description of this process, based on iteration of
          $T$, is  given elsewhere \cite{Benioff1}. 
          \begin{figure}
          \vspace{30mm}
          \caption{Initial and Final States for Counting GQTM for the First
          $2^{n}$ Binary Numbers.  All lattice qubits are in state $\vert
          0\rangle$ except those at sites $0$ and $n+1$ which are in state
          $\vert 2\rangle$.  The initial and final head states are shown as
          wave packets with internal head states $\vert 0\rangle$ and
          $\vert 1\rangle$ to the left and right respectively.}
\label{fig:fig1}
          \end{figure}
          The potential distribution associated with this example is
          obtained from a function ${\cal R}_{N}$ from the integers to
          $\{0,1\}$ such that ${\cal R}_{N}(m)=1$ if and only if the $5th$
          term of $T$, Eq. \ref{Tex} is active in the transition $T\vert
          m,i\rangle \rightarrow \vert m+1,i\rangle$ (i.e a read-1 step). 
          The potential distribution, obtained from ${\cal R}_{N}$ and Eq.
          \ref{vjj} \cite{Benioff2} is given by
          \begin{equation}
          \langle m^{\prime}\vert V_{i}\vert m\rangle = K(1-\gamma)[{\cal
          R}_{N}(m)\delta_{m^{\prime},m+1}+{\cal R}_{N}(m-
          1)\delta_{m^{\prime},m-1}]. \label{vexjj}
          \end{equation}
          Note that for each component in the initial head wave packet
          state of Figure \ref{fig:fig1} there is a different function ${\cal
          R}_{N}$.  This difference is ignored here because the functions
          differ only by a translation and describe the same potential
          distribution \cite{Benioff1}.

          It turns out  that the distributions of $1s$ and $0s$ in ${\cal
          R}_{N}$, exclusive of the $0$ tails, can be represented as an
          initial segment of length $2^{N}$ of the heirarchical sequence
          \cite{HuKe,IgTu}. If one defines the integers $\underline{m}$ as
          finite strings of $0s$ and $1s$ by $\underline{m} =01^{m}0^{m+1}$
          where $1^{m}$ and $0^{m+1}$ denote strings of $m$ $1s$ and $m+1$
          $0s$, then (except for the $0$ tails) ${\cal R}_{N}=
          \underline{0},\underline{1},\underline{0},\underline{2},
          \underline{0},\underline{1},\underline{0},\underline{3},\cdots ,
          \underline{N}$.  This definition of ${\cal R}_{N}$ is based on the
initial state $\vert 1,N+1,S\rangle$ with $\vert S\rangle$ as in Figure 
\ref{fig:fig1} chosen to be the path origin.
          
  This sequence is an initial segment of length $2^{N}$ of the heirarchical
          sequence in the underlined numbers.  Based on this it can be
          shown that ${\cal R}_{N}$ can be generated by a pair of recursion
          relations given by \cite{Benioff1,Benioff2}
          \begin{equation}
          {\cal R}_{n}={\cal S}_{n-1}\underline{n};\: {\cal S}_{n}={\cal
          R}_{n}{\cal S}_{n-1}. \label{calRrecur}
          \end{equation}
          for $n=1,2,\cdots ,N$ with ${\cal S}_{0}=\underline{0}$.  

          Since the recursion relations generate the potential distribution
          it can be seen that the transfer matrix $Z_{N}$ is given by a
          pair of recursion relations obtained from Eq.\ref{calRrecur} as 
          \begin{eqnarray}
          Z_{n} & = & W_{n}X_{n-1} \nonumber \\
          X_{n} & = & X_{n-1}W_{n}X_{n-1} \label{Zrecur}
          \end{eqnarray}
          with $X_{0}|_{11} =\mathrm{e}^{2\mathrm{i}k}$ and $X_{0}|_{12}=0$
          for $n=1,2\cdots ,N$. $W_{n}$ is the transfer matrix for the
          potential corresponding to the sequence $\underline{n}
          =01^{n}0^{n+1}$.  Note that the order of matrix multiplication is
          the inverse of the order in which terms appear in the sequence of
          Eq. \ref{calRrecur}.  

          The main advantage of these recursion relations is that only
          polynomially many matrix multiplications are needed to obtain the
          elements of the matrix $Z_{N}$.  If $Z_{N}$ is obtained from the
          matrices associated with each potential barrier in the
          distribution, then exponentially many matrix multiplications are
          required.

          From Eq. \ref{Zrecur} one sees that the only matrices needed in
          explicit form (other than $X_{0}$) are the $W_{m}$ for $m \leq
          N$.  These are given explicitly by \cite{Benioff2}
          \begin{eqnarray}
          W_{m}|_{11} & = & \frac{\mathrm{e}^{\mathrm{i}k(m+2)}} 
	  {2\mathrm{i}\gamma \sin{k}\sin{l}} [\mathrm{e}^{2\mathrm{i}k} 
	  \sin{lm}-2\gamma \mathrm{e}^{\mathrm{i}k}\sin{l(m-1)} 
	  +\gamma^{2}\sin{l(m-2)}] \nonumber \\
          W_{m}|_{12} & = & \frac{\mathrm{e}^{\mathrm{i}km}} 
	  {2\mathrm{i}\gamma \sin{k}\sin{l}}[\sin{lm}-2\gamma 
	  \sin{l(m-1)}\cos{k} +\gamma^{2}\sin{l(m-2)}].  \label{Zmat}
          \end{eqnarray}
          Here the momenta $l,k$, which refer to the core and potential
          free regions, are related by $ \cos k =\gamma \cos l$.  Note that
          $W_{m}|_{11}=W^{*}_{m}|_{22},\: W_{m}|_{21}=W^{*}_{m}|_{12}$ and
          $W_{m}$ is unimodular.

          These equations are valid for the unbound region $E\geq V$.  The
          equivalent equations for the bound region are obtained by
          replacing $\sin$ everywhere in Eqs. \ref{Zmat} by
          $\mathrm{i}\sinh$, the hyperbolic sine function.  Also $\cos
          k=\gamma \cosh l$.

          \section{Landauer Resistance, Band Spectra}
          \label{BSLR}
          Calculations are made here for the LR and band spectra for
          several different values of $\gamma$ and $N$ and momentum regions
          for the counting GQTM.  The band spectra refer to the associated
          periodic system which is a crystal whose unit cell is the
          potential distribution corresponding to enumeration of the first
          $2^{N}$ integers.  The momentum regions include the bound and
          unbound state regions.  There are $\gamma$ dependent lower bounds
          on the values of $|k|$ below which there are no bands and the LR
          is very large \cite{Benioff2}; these regions of $k$ values are
          excluded from the calculations.

          Each figure gives a plot of the Log (base 10) of LR as a function
          of momentum $k$ for chosen values of $N$ and $\gamma$ and a
          region of $k$ values.  The $\log LR$ is plotted instead of the LR
          because the LR fluctuates rapidly and violently over many orders
          of magnitude and the interest here is in regions of appreciable
          transmission (LR values of order unity).  The band and gap
          spectra for the same parameters is also included so that one can
          compare the LR fluctuations to the band spectra.  The number of
          momentum values used to generate the curves in each figure ranges
          from about 3,000  to 8,000.

          As both the LR and band spectra are extremely sensitive to values
          of $\gamma$, values that are reasonable physically should be
          chosen. This is done \cite{Benioff2} by recalling that if $T$ is
          the bilateral shift on the qubit lattice, the Feynman
          Hamiltonian. Eq. \ref{ham} is the symmetrized lattice equivalent
          of $(-\hbar^{2}/2m)d^{2}/dx^{2}$.  From this and $V=2K\delta$
          where $\delta=1-\gamma$, one obtains $\delta
          =Vm\Delta^{2}/\hbar^{2}$ where $\Delta$ is the qubit lattice
          spacing.  For electron systems, $m$ is of the order of the
          electron mass, $V$ is a few electron volts, and $\Delta$ is
          measured in Angstroms.  Taking $m$ equal to 2 electron masses,
          $\Delta = 1 \AA$, and $V=2ev$ gives $\delta \simeq 0.001$ or
          $\gamma \simeq 0.999$.

          Using this value of $\gamma$ the $\log LR$ and band spectra for
          the associated periodic system have been calculated for $N=10$
          for several momentum or energy regions.  The initial qubit state
          is shown in Figure \ref{fig:fig1}.  The results are shown in Figures
          \ref{fig:fig2} and \ref{fig:fig3} for momentum ranges of $0.0223 \leq k
          \leq 0.10$ and $0.095\leq k\leq 0.220$.  The bound state region
          extends up to $k=0.0447$ with the unbound region extending up and
          beyond the upper limit.  In this and all succeeding figures the
          band spectrum is shown as a band with upper or lower flat line
          segments denoting bands or gaps respectively.  Very narrow gaps
          (or bands) are shown as downward (or upward) pointing spikes.
          \begin{figure}
                 \vspace{30mm}
          \caption{The Log (base 10) of the Landauer Resistance Plotted as
          a Function of the Momentum for $N=10$ and $\gamma =0.999$ for a
          Momentum Range $0.0223\leq k\leq 0.10$. The energy band spectrum
          for the associated periodic system is shown at the bottom of the
          figure where upper horizontal line segments correspond to energy
          bands and lower horizontal segments correspond to energy gaps. 
          Very short bands or gaps appear as points.  Band-gap edges show
          as vertical lines. The ordinal placement of the band spectrum, at
          the bottom of the figure, is done for convenience only.}
          \label{fig:fig2}
          \end{figure}
          \begin{figure}
          \vspace{30mm}
          \caption{The Log (base 10) of the Landauer Resistance Plotted as
          a Function of the Momentum for $N=10$ and $\gamma =0.999$ for a
          Momentum Range $0.095\leq k\leq 0.220$. The energy band spectrum
          for the associated periodic system is shown at the bottom of the
          figure.  Additional details are given in the caption for Figure
          2.} \label{fig:fig3}
          \end{figure}
          The results in the two figures show that the characteristics of
          the band spectra and $\log LR$ found for $k\leq 0.0447$
          \cite{Benioff2}, extend smoothly into the region where $E>V$. 
          Most of the region is occupied by bands with most gaps being
          quite narrow.  A few wider gaps, such as those at $k=0.033,\;
          0.077,\; 0.149$ and especially at $k=0.054,\; 0.101,\; 0.198$ are
          present.      

          The $\log LR$ can be characterized as a downward trending band of
          fluctuating values from an average value of about $-1,\;
          (LR=0.1)$  at the lower end of the $k$ region to about $-2,\;
          (LR=0.01$) at the upper end of Figure \ref{fig:fig2} to a value of
          about $-3.5$ at the upper end of Figure \ref{fig:fig3}.  Several
          peaks extending to higher values (up to $LR \simeq 10$) project
          out of the band.  

          The peaks in $\log LR$ appear to be associated with gaps, with
          the higher peaks associated with wider gaps.  This effect
          \cite{Benioff2} was also found for calculations with the Kronig-
          Penney model with potential distributions corresponding to the
          Fibonacci and Thue-Morse substitution sequences \cite{Roy1,Roy2}. 
          The minima in $\log LR$ which show as downward pointing spikes
          appear to be associated also with narrow gaps.  Additional
          details on this are given elsewhere\cite{Benioff2}. 



          Results of calculations of the band spectrum and $\log LR$ for a
          larger value of $N=18$, corresponding to a potential distribution
          with $2^{17}$ individual potential barriers, are shown in the next
          two figures for the same two momentum ranges that were used
          in Figures \ref{fig:fig2} and \ref{fig:fig3}.  The same value of $\gamma
          =0.999$ was used.  The band spectra are much more finely divided
          with each relatively broad band for $N=10$ divided into a great
          number of very narrow bands with intervening narrow gaps.  The
          fraction of the momentum regions occupied by bands is smaller
          than that for $N=10$.  The LR fluctuates very rapidly over many
          orders of magnitude between low values less than $10^{-4}$ to
          high values greater than $10^{20}$.  For much of the momentum
          region $\log LR$ occupies a band (the black region in the
          figures) which trends down from a value around $0\; (LR=1)$ at
          the low momentum end of Figure \ref{fig:fig4} to a value around $-3\;
          (LR=10^{-3})$ at the high momentum end of Figure \ref{fig:fig5}.  The
          trend, which represents a kind of average, is in the right
          direction as the LR should decrease as the energy is increased.
          \begin{figure}
          \vspace{30mm}
          \caption{The Log (base 10) of the Landauer Resistance Plotted as
          a Function of the Momentum for $N=18$ and $\gamma =0.999$ for a
          Momentum Range $0.0223\leq k\leq 0.10$. The energy band spectrum
          for the associated periodic system is shown at the bottom of the
          figure.  Additional details are given in the caption for Figure
         2.} \label{fig:fig4}
          \end{figure}
          \begin{figure}
          \vspace{30mm}
          \caption{The Log (base 10) of the Landauer Resistance Plotted as
          a Function of the Momentum for $N=18$ and $\gamma =0.999$ for a
          Momentum Range $0.095\leq k\leq 0.22$. The energy band spectrum
          for the associated periodic system is shown at the bottom of the
          figure.  Additional details are given in the caption for Figure
          2.} \label{fig:fig5}
          \end{figure}

          Probably the most remarkable aspect is that the transmission
          probability is high ($>50-99\%$) on average, yet it is filled
          with numerous holes as small regions of very low or negligible
          transmission.   This is shown by the numerous excursions of $\log
          LR$ to high values $>4$ from low average values around $0$ to $-
          2$ and lower in Figures \ref{fig:fig4} and \ref{fig:fig5}.  It is
          noteworthy that this effect appears to be independent of whether
          $E\leq V$ or $E>V$ ($E=V$ at $k=\arccos{\gamma} = 0.0447$).  Also
          there is a correlation between the widths of regions of low
          transmission and the associated band spectrum gap widths in that
          wider gaps are correlated with wider regions of low transmission. 
          Examples are the relatively wide band gaps at $k=0.054,\;
          0.100,\; 0.198$. For these gaps the widths of the regions of very
          low transmission (high values of $\log LR$) are relatively wide. 

          Calculations made for the same values of $N=18$ and $\gamma
          =0.999$ for higher values of the momentum, up to $k=0.5$, show
          that the pattern described above continues.  The "black" band
          containing much of the fluctuating values of $\log LR$ continues
          a slow downward trend to $\approx -4$ at the upper end of the
          region.  Numerous narow spikes to high values are present.  In
          some narrow momentum regions the spikes are so dense that high
          resolution calculations are needed for their individual
          resolution. 

          In order to investigate the dependence of the LR and band spectra
          on $\gamma$, calculations were made for a value of $\delta $
          larger by a factor of $10$, i.e. $\gamma =0.99$, and $N=10$.  The
          results for the energy regions $0.069\leq k\leq 0.220$ and
          $0.20\leq k\leq 0.42$ are shown in the next two figures.
          \begin{figure}
          \vspace{30mm}
          \caption{The Log (base 10) of the Landauer Resistance Plotted as
          a Function of the Momentum for $N=10$ and $\gamma =0.99$ for a
          Momentum Range $0.069\leq k\leq 0.22$. The energy band spectrum
          for the associated periodic system is shown at the bottom of the
          figure.  Additional details are given in the caption for Figure
          2.} \label{fig:fig6}
          \end{figure}
          \begin{figure}
          \vspace{30mm}
          \caption{The Log (base 10) of the Landauer Resistance Plotted as
          a Function of the Momentum for $N=10$ and $\gamma =0.99$ for a
          Momentum Range $0.20\leq k\leq 0.42$. The energy band spectrum
          for the associated periodic system is shown at the bottom of the
          figure.  Additional details are given in the caption for Figure
          2.} \label{fig:fig7}
          \end{figure}

          The band spectra show the presence of a number of wide gaps. 
          Also most of the bands are quite narrow.  Comparison with the
          band spectra of Figs. \ref{fig:fig2} and \ref{fig:fig3}, which are 
	  for the same value of $N$  but $\delta$ smaller by a factor of $10$,
          show that the fraction of the momentum regions occupied by bands
          is much lower and gap widths are wider.  $\log LR$ shows a band
          of fluctuating values decreasing slowly from an average value of
          $\approx 1 (LR=10)$ at the low end of the momentum region in
          Figure \ref{fig:fig6} to a value of $\approx -2 (LR=0.01)$ at the
          upper end of the momentum region in Figure \ref{fig:fig7}.    

          Superimposed on this are peaks in the values of $\log LR$ which
          extend up to $18$ or more.  Each peak appears to occupy the same
          momentum region as the gap, with the peak height correlated with
          the gap width in that wider gaps are associated with higher
          peaks.  This is especially clear in Figure \ref{fig:fig6}.  The
          distribution of peak heights in the figure also shows a
          heirarchical relationship similar to that shown by the potential
          distribution of Eqs. \ref{vexjj} and \ref{calRrecur}.  That is,
          the region between the two largest peaks is divided in two by a
          peak of lesser height.  Each subregion is divided in two by an
          even smaller peak.  This division into two subregions by peaks of
          decreasing height continues down to the smallest peaks associated
          with the small gaps and which form the band of fluctuating
          values.  This regularity is less evident in the momentum region
          shown in Figure \ref{fig:fig7} for which the fluctuations in $\log
          LR$ are more chaotic.  Since the value of $k=0.142$ for which
          $E=V$ occurs in the middle of Figure \ref{fig:fig6}, it appears that
          the regularity of the peak distribution is not dependent on
          whether the momentum is in the bound or unbound region. 

          In order to investigate in more detail the relation between peaks
          in $\log LR$ and bands and gaps, a high resolution calculation
          for $\gamma=0.99,\; N=10$ was done for the small momentum region
          $0.21\leq k\leq 0.23$.  The results are shown in the next figure.
          The results show that each peak in $\log LR$ is
          defined by sharp minima which are associated with relatively
          narrow gaps.  More exactly the minima seem to be associated with
          the band-above-gap edges of the narrower gaps.  In this fashion
          the momentum region of each peak includes one band and one gap. 
          Each gap has a sharp minima associated with its upper edge except
          for the wider gaps.  This appears to result from the fact that in
          general the depth of the minima in $\log LR$ decreases with
          increasing width of the gap at whose edge it is located until it
          disappears entirely for sufficiently wide gaps.  For these wider
          gaps the momentum region of the associated peak includes two
          bands and two gaps.  This effect, which was noted before
          \cite{Benioff2}, is seen for the wide gaps centered at
          $k=0.216,\; 0.2185,\; 0.221,\; 0.2267$. 
          \begin{figure}
          \vspace{30mm}
          \caption{The Log (base 10) of the Landauer Resistance Plotted as
          a Function of the Momentum for $N=10$ and $\gamma =0.99$ for a
          Small Momentum Range $0.21\leq k\leq 0.23$. The energy band
          spectrum for the associated periodic system is shown at the
          bottom of the figure.  Additional details are given in the
          caption for Figure 2.} \label{fig:fig8}
          \end{figure}

          Increasing $\delta$ by another factor of $10$ to $0.1$ with
          $\gamma =0.9$ gives the results shown in the next two figures
          for the momentum regions $0.18\leq k\leq 0.52$ and
          $0.5\leq k\leq 1.0$ respectively.  The calculations are for
          $N=8$.  Since $k=0.45$ for $E=V$, the energy region used in
          Figure \ref{fig:fig9} is mostly the bound state region.  That used in
          Figure \ref{fig:fig10} includes part of the unbound state region.  
          \begin{figure}
          \vspace{30mm}
          \caption{The Log (base 10) of the Landauer Resistance Plotted as
          a Function of the Momentum for $N=8$ and $\gamma =0.9$ for a
          Momentum Range $0.18\leq k\leq 0.52$. The energy band spectrum
          for the associated periodic system is shown at the bottom of the
          figure.  Additional details are given in the caption for Figure
          2.} \label{fig:fig9}
          \end{figure}
          \begin{figure}
          \vspace{30mm}
          \caption{The Log (base 10) of the Landauer Resistance Plotted as
          a Function of the Momentum for $N=8$ and $\gamma =0.9$ for a
          Momentum Range $0.5\leq k\leq 1.0$. The energy band spectrum for
          the associated periodic system is shown at the bottom of the
          figure.  Additional details are given in the caption for Figure
          2.} \label{fig:fig10}
          \end{figure}
          The results in the figures show that the potential is high enough
          (10 times the height of that in Figures \ref{fig:fig6} and
          \ref{fig:fig7}) so that definite differences in the band spectra and
          $\log LR$ are evident for the two regions.  In the bound state
          region (Fig. \ref{fig:fig9}) the bands are extremely narrow and are
          few and widely separated with many large gaps present. The values
          of $\log LR$ show the the LR is very high for most of the region,
          except for a few very narrow spike minima to values around $1,\;
          (\log LR =0)$.  The spike minima occur at the same energies as
          the bands.  

          In the unbound region (Fig. \ref{fig:fig10}) the bands are much more
          numerous and individual bands are wider than in the bound state
          region.  The LR is much lower on average, and there are extended
          regions ($0.59\leq k\leq 0.62$ and $0.70\leq k\leq 0.81$) for
          which $LR\simeq 1$.  To summarize, the results for these figures
          show that in the bound state region there is almost no
          transmission through the 128 potentials except for a few
          extremely narrow regions at which bands are located.  In the
          unbound state region there are wide regions of good transmission
          of $\simeq 50\%$ or more. Energy bands are more numerous and are
          wider.

          \section{discussion}
          \label{disc}
          The values of $\gamma$ and $N$ for which results of calculations
          of band spectra and $\log LR$ have been shown are quite limited.
          These limitations are imposed by the requirement that the
          calcuations of the band spectra and LR be reasonably reliable. 
          This is not trivial because fluctuations in the values of the
          matrix elements of $Z_{N}$ used to calcuate the LR and the band
          spectra become extremely rapid and extreme at higher values of
          $N$ and lower values of $\gamma$. For example, even for $N=18$
          and $\gamma=0.999$ or $N=8$ and $\gamma=0.9$, $TrZ_{N}$ and the
          LR fluctuate rapidly between $\pm10^{M}$ with $M$ taking values
          of 200-300 or more for some values of $k$. Also these
          fluctuations occur over extremely small intervals of $k$.

          In essence limitations on the values of $N$ and $\gamma$ are
          imposed by (classical) computer program limitations on the
          magnitude and number of significant figures allowed in the
          computations.  Modifications of the programs to allow
          calculations over a larger range of $N$ and $\gamma$ values are
          not warranted as they are complex and at most give one a slight
          extension of the range of acceptable values of $N$ and $\gamma$.

          In many ways the results obtained here, which are similar to
          those obtained by Roy and Khan \cite{Roy1,Roy2}, are unexpected. 
          The LR fluctuates rapidly with many spikes down to extremely low
          values over the momentum range examined.  Except for the highest
          potential considered (i.e. $\gamma=0.9$) there is not much
          difference between the bound and unbound regions as far as the
          band spectra or $\log LR$ are concerned.  The transmission, even
          in the bound region for $N=18$ and $\gamma =0.999$, is high
          ($\simeq 50\%$ or more) on average although there are a great
          many holes of extremely low transmission.

          These results need explaining since one would expect little or no
          transmission in the bound region, because of decay in the
          transmission amplitude resulting from tunnelling effects and
          localization \cite{Landauer}.  This is especially the case for
          $N=18$ with more than $128,000$ potential barriers present.  Also
          one would expect almost complete transmission in the unbound
          regions with few or no regions of low transmission.  Neither of
          these results occur.

          It is suggested here that the calculated results are best
          understood in terms of coherent superposition of the many
          reflected waves moving backwards through the  potential
          distribution. At some values of the momenta the reflected waves
          interfere constructively, which greatly increases the amplitude
          of the overall reflected component and decreases the amplitude to
          the transmitted component.  At other momentum values the
          reflected waves interfere destructively. In this case the
          amplitude of the reflected component is small and the amplitude
          of the transmitted component is high. \footnote{Exactly the same
          argument applies {\em mutatis mutandis}. Constructive or
          destructive interference of the transmitted components decreases
          or increases the amplitude of the reflected component emerging
          from the distribution.  From the viewpoint of a computation one
          is more interested in the amplitude of the transmitted component
          than in the reflected one.}  Note that if $B$ and $F$ are the
          complex valued reflection and  transmission coefficients
          respectively for the potential distribution,Eqs. \ref{PsiIMF} and
          \ref{Zdef}, then normalization requires that $|B|^{2}+|F|^{2}=1$. 

          The number of component waves which add coherently should be very
          large. Since each barrier in the distribution generates
          transmitted and reflected components from any impinging wave,
          whether it is the input wave, or waves reflected or transmitted
          from other barriers, the number of multiple reflections or
          transmissions or combinations of the two becomes extremely large,
          especially in a distribution with many barriers.  All the
          components moving in the same direction add coherently.

          This coherent superposition is expected to depend sensitively on
          all parameters involved. It also should be very dependent on the
          fact that the potential distribution is not random; in the
          example considered here the distribution is quasiperiodic
          \cite{FuO,DiVSt} and is heirarchical.  Intuitively if coherent
          superposition is the dominant effect, one would expect that for
          fixed potential height (fixed $\gamma$), the fluctuation rate of
          the Landauer Resistance and the number of bands and gaps to
          increase with increasing $N$. The reason is that the number of
          individual potential barriers present increases exponentially
          with increasing $N$. The corresponding increase in the number of
          reflected and transmitted component waves which combine
          coherently means that momentum regions in which backward moving
          reflections interfere destructively to give high transmission
          become much narrower and broad bands split into many narrower
          bands with intervening gaps.

          In a similar fashion momentum regions in which backward moving
          reflections combine constructively to give very low transmission
          also become narrower.  This narrowing and splitting of regions
          with constructive or destructive interference results in a large
          increase in the extent and rapidity of fluctuations in the LR and
          splitting of the energy bands.

          The importance of coherent effects in the transmission and
          reflection of a state through the potential distribution is
          demonstrated clearly by consideration of the transmission and
          reflection for a single potential barrier. For motion in the
          potential distribution corresponding to iteration of all binary
          numbers of length $\leq N$ the complex transmission (F) and
          reflection (B) coefficients are obtained from Eqs. \ref{Zdef} and
          \ref{Zprop} and the normalization $|F|^{2}+|B|^{2}=1$ as
          \begin{equation}
          F=\frac{1}{Z_{N}|_{22}},\; B=-\frac{Z_{N}|_{12}}{Z_{N}|_{22}}.
          \label{FBdef}
          \end{equation}

          For a single barrier corresponding to $m$ read 1 steps these
          equations hold with $Z_{N}$ replaced by $W_{m}$, Eq. \ref{Zmat}.
          The momentum dependence of the magnitudes and phases of F and B
          for $m=10$ and $\gamma =0.999$ for a momentum range of $0\leq
          k\leq 0.5$ are shown in the next two figures. The magnitude
          values for the reflection coefficient $B$ for $k\geq 0.3$ have
          been multiplied by a factor of 10 to illustrate the nodal
          structure of $B$ more clearly. The range of the phases in Figure
          \ref{fig:fig12} extends from $-\pi$ to $\pi$.  The large
          discontinuities in the curves for both B and F are a artifact of
          the presentation method in that $\pi +x\equiv -\pi +x$. The
          discontinuity in the phase for B at $k\simeq 0.3$ is real,
          though.  It occurs at the node point for $|B|$, Figure
          \ref{fig:fig11}.
          \begin{figure}
          \vspace{30mm}
          \caption{The magnitudes of the Reflection (B) and Transmission
          (F) Coefficients for a Single Potential Barrier Corresponding to
          $m=10$ read 1 Steps for $\gamma =0.999$ for a Momentum Range of
          $0\leq k\leq 0.5$. The values of $|B|$ for $k\geq 0.3$ were
          increased by a factor of $10$ to show more clearly the nodal
          structure.} \label{fig:fig11}
          \end{figure}
          \begin{figure}
          \vspace{30mm}
          \caption{The Phases of the Reflection (B) and Transmission (F)
          Coefficients for a Single Potential Barrier Corresponding to
          $m=10$ read 1 Steps for $\gamma =0.999$ for a Phase Range of $[-
          \pi,\;\pi]$.  The large discontinuities in the phases for $B$ and
          $F$ are a result of the figure presentation in which $-\pi$ and
          $\pi$ are not identified.} \label{fig:fig12}
          \end{figure}
          The figures show that the magnitudes of B and F tend rapidly to
          values close to $0$ and $1$. At a value of $k=0.0447$
          corresponding to $E+V$, $|F|= 0.976$ and $|B|=0.218$.  Thus at
          momenta corresponding to energies above the barrier height there
          is still appreciable reflection.  Unlike the magnitudes the
          phases of both B and F vary steadily over
          the range of momenta shown.  This is significant because it means
          that coherent additions of the many reflected and transmitted
          components in a sequence of these barriers would be expected to
          give rapidly varying overall transmission and reflection
          coefficients.  This is shown by the calculations.

          Calculations show that as $N$ is increased for constant $\gamma$
          the internodal maxima increase and the distances between nodes in
          $|B|$ decrease.  The rate of change of the phases of both F and B
          also increases and the discontinuities at the B node points
          increase.  This implies an increase in the rapidity of
          fluctuations of the LR and in the number of spectral bands and
          gaps with increasing $N$, also in agreement with the
          calculations.

          As the potential is increased ($\gamma$ decreased) but $N$ is
          fixed, the position of the first node in $|B|$ moves out to
          higher values of $k$.  Also the internodal maxima increase. 
          However the rates of change of the phases for both B and F
          decrease out to the first node as the potential height increases.
          For higher values of $k$ the phase slopes appear to be roughly
          independent of the potential height.  It is of interest to note
          that the position of the first node at $k=0.59$ for
          $\gamma=0.9,\; N=8$ is close to the value of $k$ at which the LR
          and band spectra change character (See discussion of Figures
          \ref{fig:fig9} and \ref{fig:fig10}).

          The curves of Figure \ref{fig:fig11} can be used to show the
          importance of coherent effects.  Consider for example the
          potential distribution for $N=18$ and $\gamma =0.999$ used to
          calculate the LR and band spectra shown in Figures \ref{fig:fig4} and
          \ref{fig:fig5}.  The properties of the heirarchical distribution are
          such that for $m=1,2,\cdots ,N-1$ there are $2^{N-m-1}$ potential
          barriers of width corresponding to m read 1 steps in the
          distribution.  For the example at hand there are $128$ barriers
          for $m=10$.  If the reflections and transmissions from these
          barriers combine incoherently, then the probability of
          transmission through just these barriers and ignoring the others
          in the distribution is given by $(1-|B(k)|^{2})^{128}$ where the
          $k$ dependence is shown explicitly.  It is clear from the $k$
          dependence of $|B(k)|$ shown in Figure \ref{fig:fig12} and the LR
          shown in Figures \ref{fig:fig4} and \ref{fig:fig5} that the $k$
          dependence of the probability of transmission is completely
          different from the incoherent prediction.  Inclusion of the
          effects of the other barriers for different $m$ in the incoherent
          prediction whould make the disagreement even more extreme.

          The coherent effects implied in the results for the LR and band
          spectra give a good illustration of nonlocality in quantum
          mechanics. Consider a value of $k$ at which the reflection
          components interfere destructively as they move backwards along
          the path. This means the amplitude of the overall reflection
          coefficient for the reflected wave emerging from the distribution
          is a (local) minimum and the amplitude for the overall emerging
          transmitted wave is a maximum. This is a nonlocal effect since
          much of what happens to the reflected components  occurs at path
          locations distant from those of the transmitted components. The
          effect is quite pronounced since it extends over the whole
          potential distribution.  In the case of $N=18$ this effect
          extends over $10^{6}$ path sites. The distance over which this
          effect occurs grows exponentially with $N$ as the number of path
          sites in the potential distribution is about equal to $2^{N+2}$.

          Is is to be emphasized that this effect is quite general and is
          not restricted to the specific distribution considered here.  It
          is shown, for instance, in the graphs of the LR versus energy for
          different potential distributions  obtained by Roy and Kahn
          \cite{Roy1,Roy2}. These effects also play an important role in
          most work on transmission of electrons through nonrandom
          potential distributions.

          \ack{This work is supported by the U.S. Department of Energy,
          Nuclear Physics Division, under contract W-31-109-ENG-38.}

          
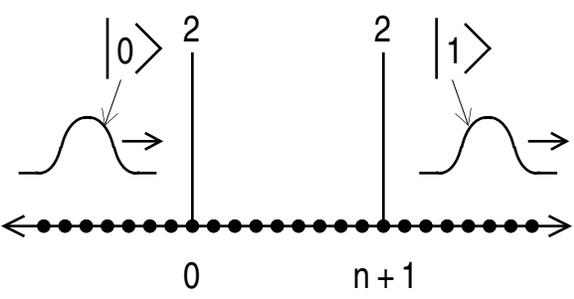
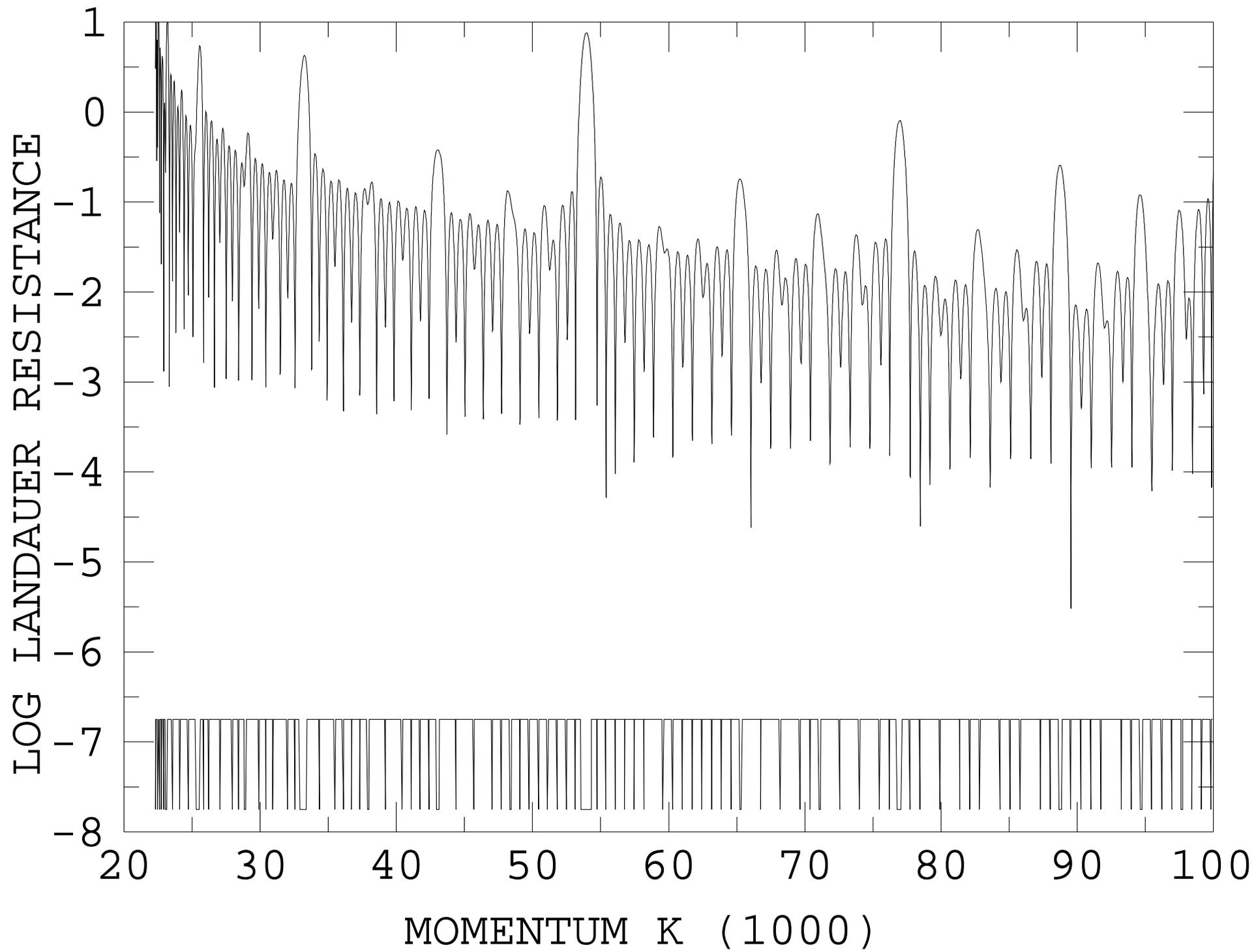
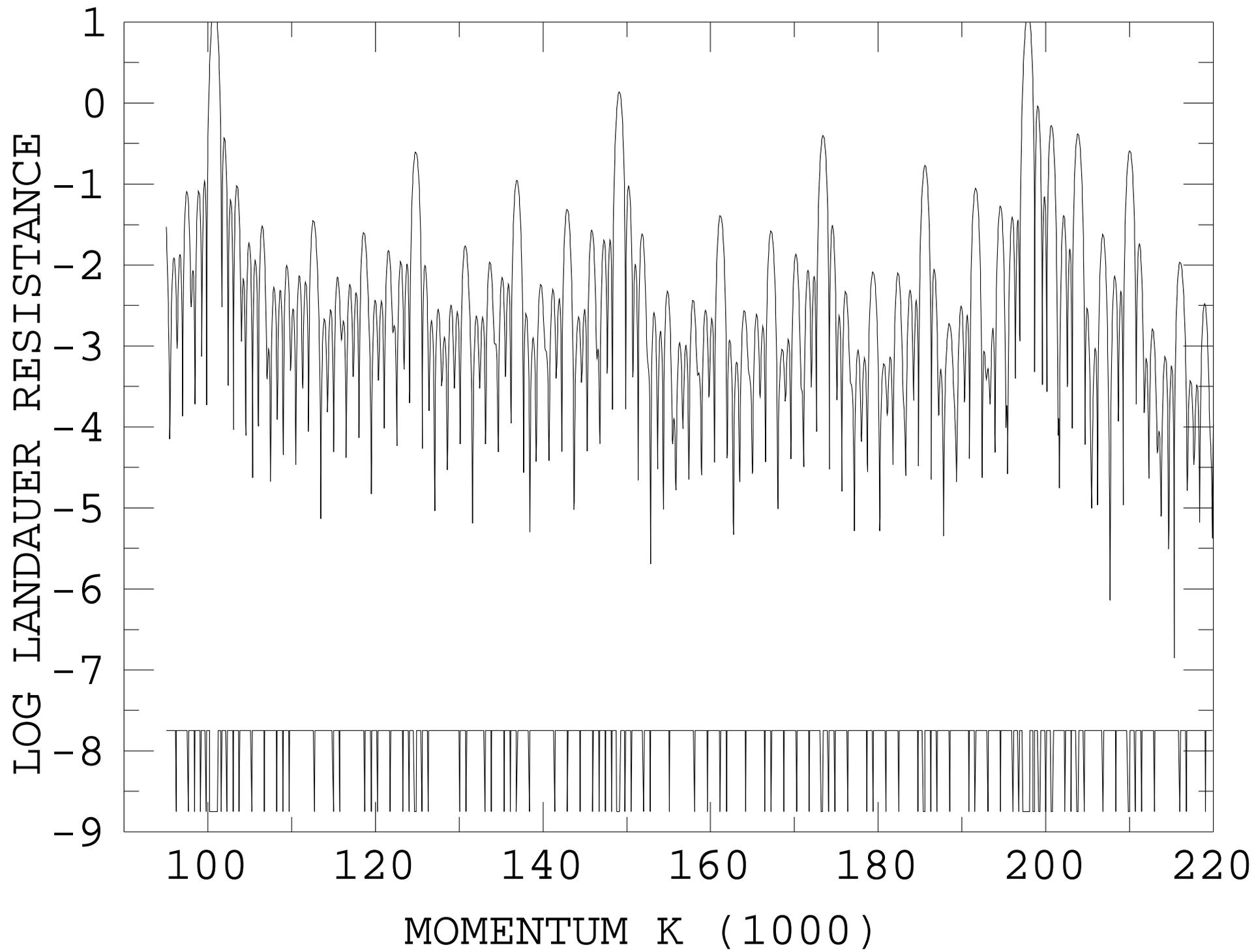
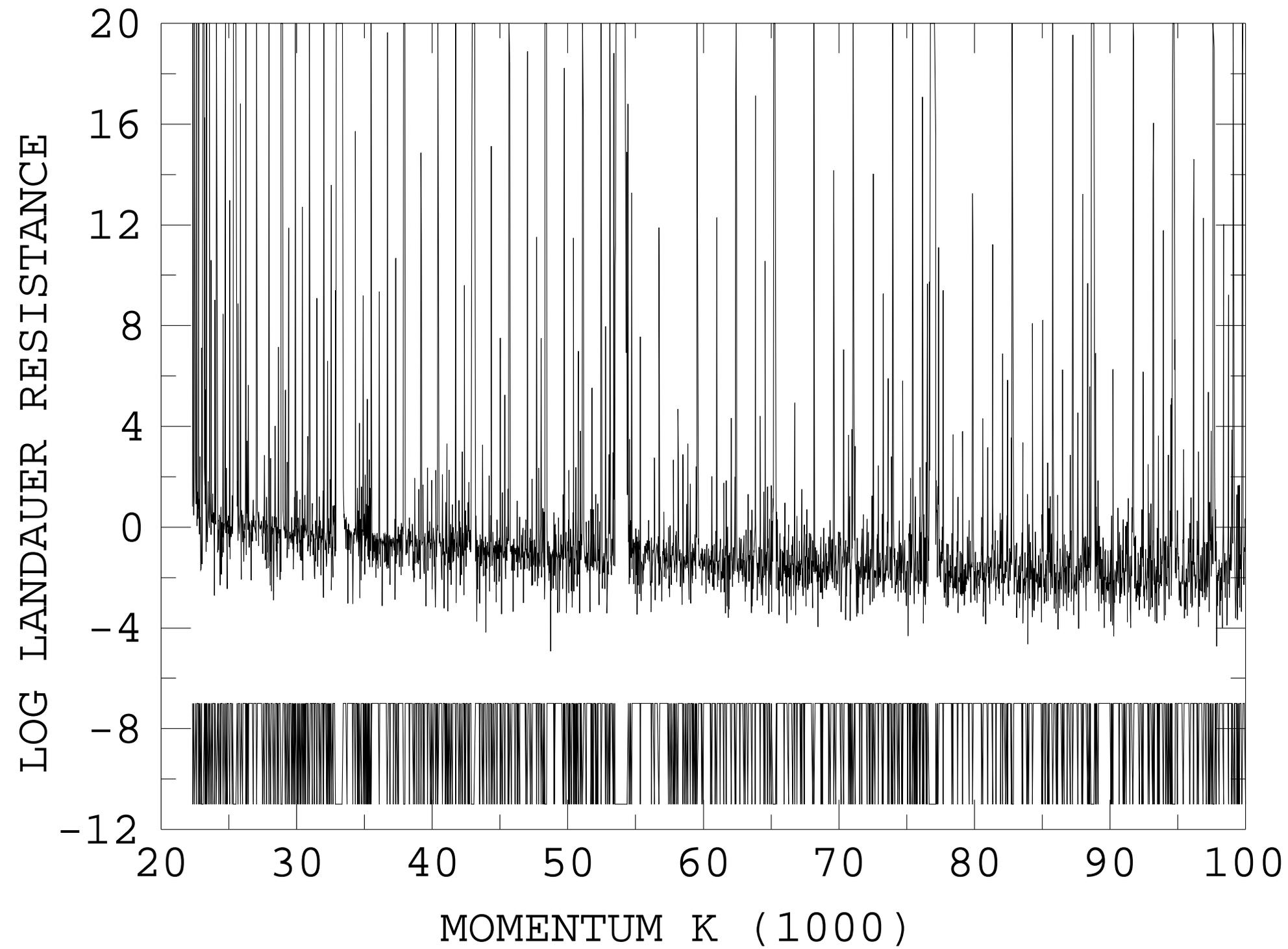
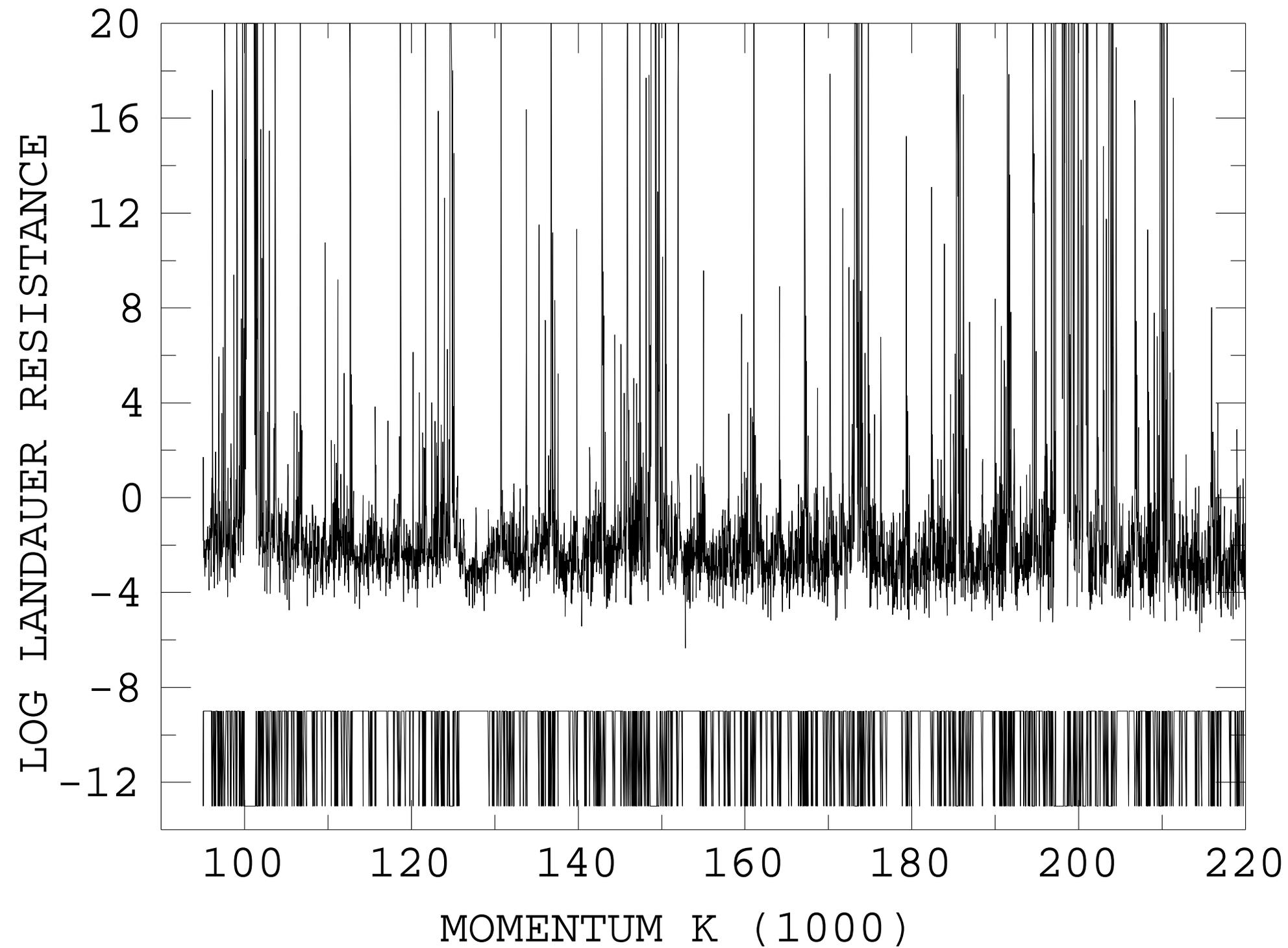
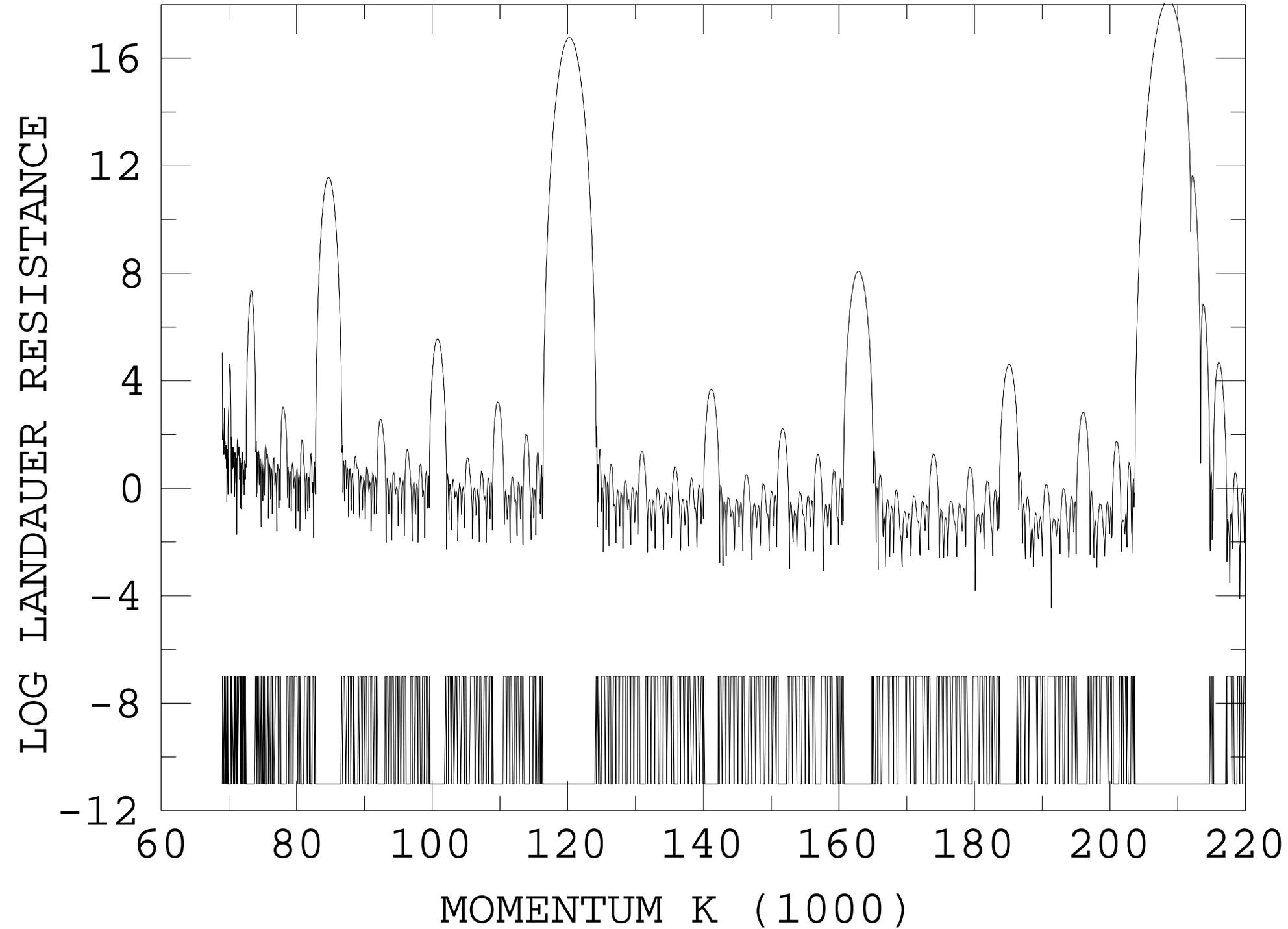
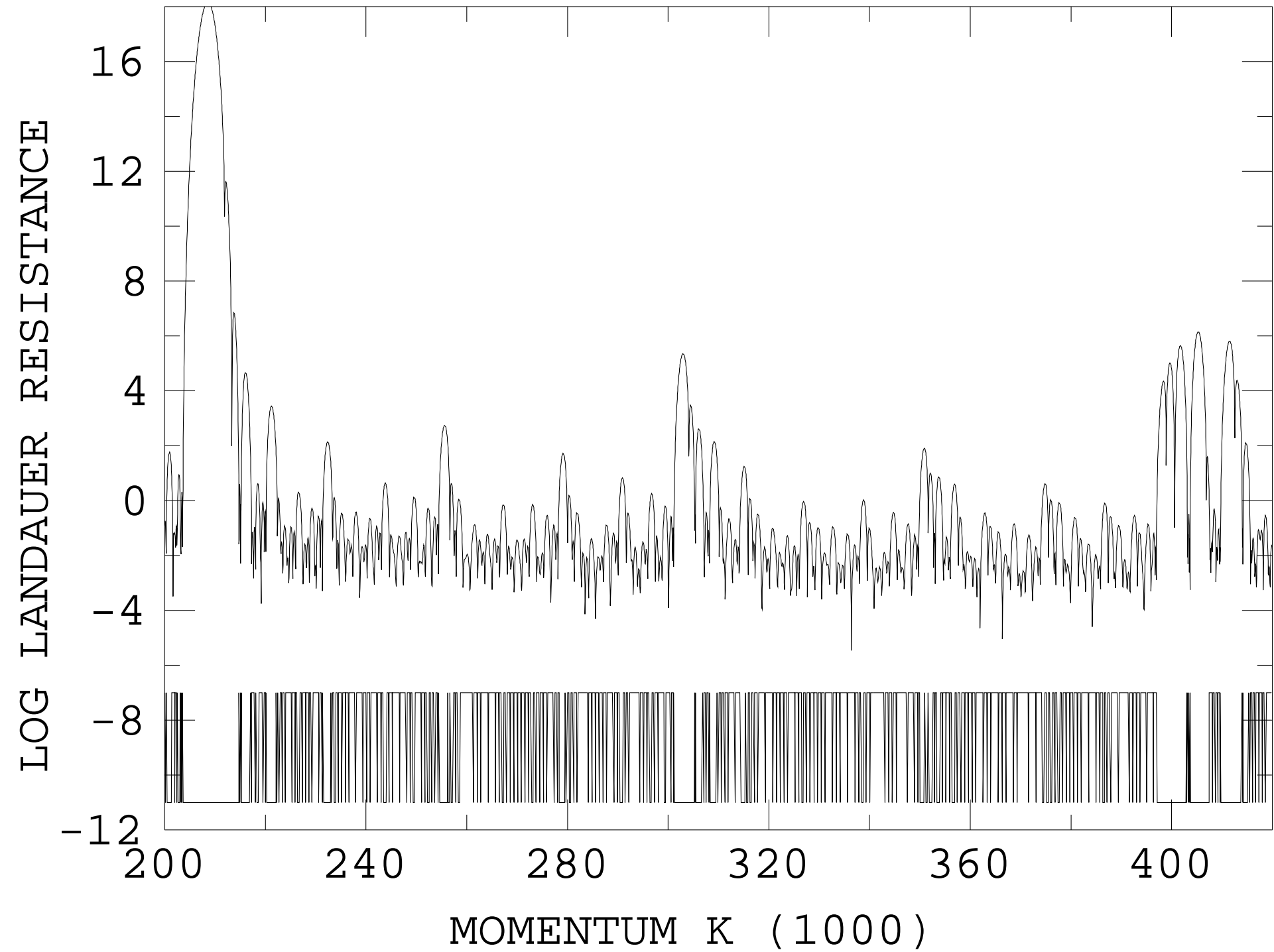
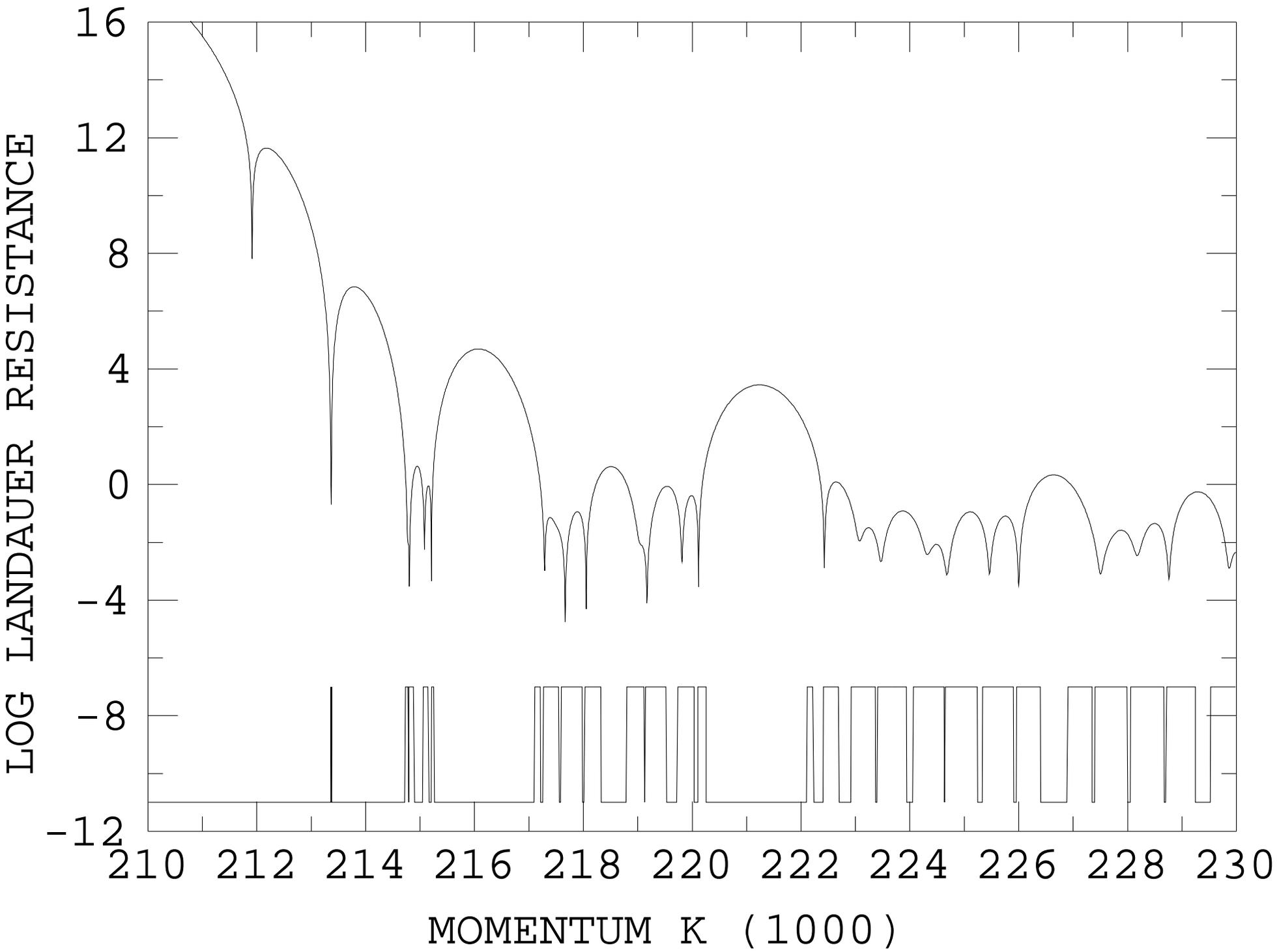
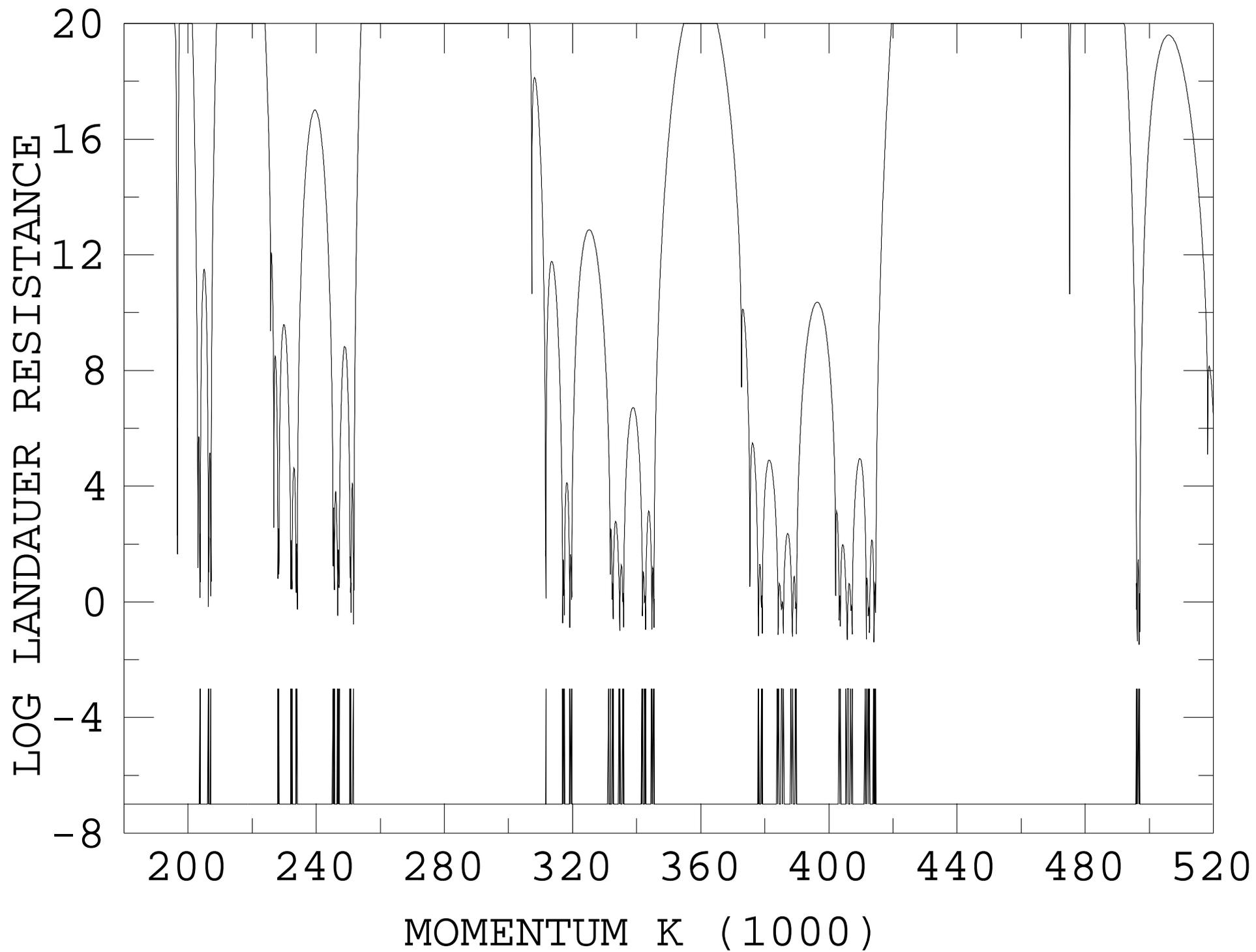
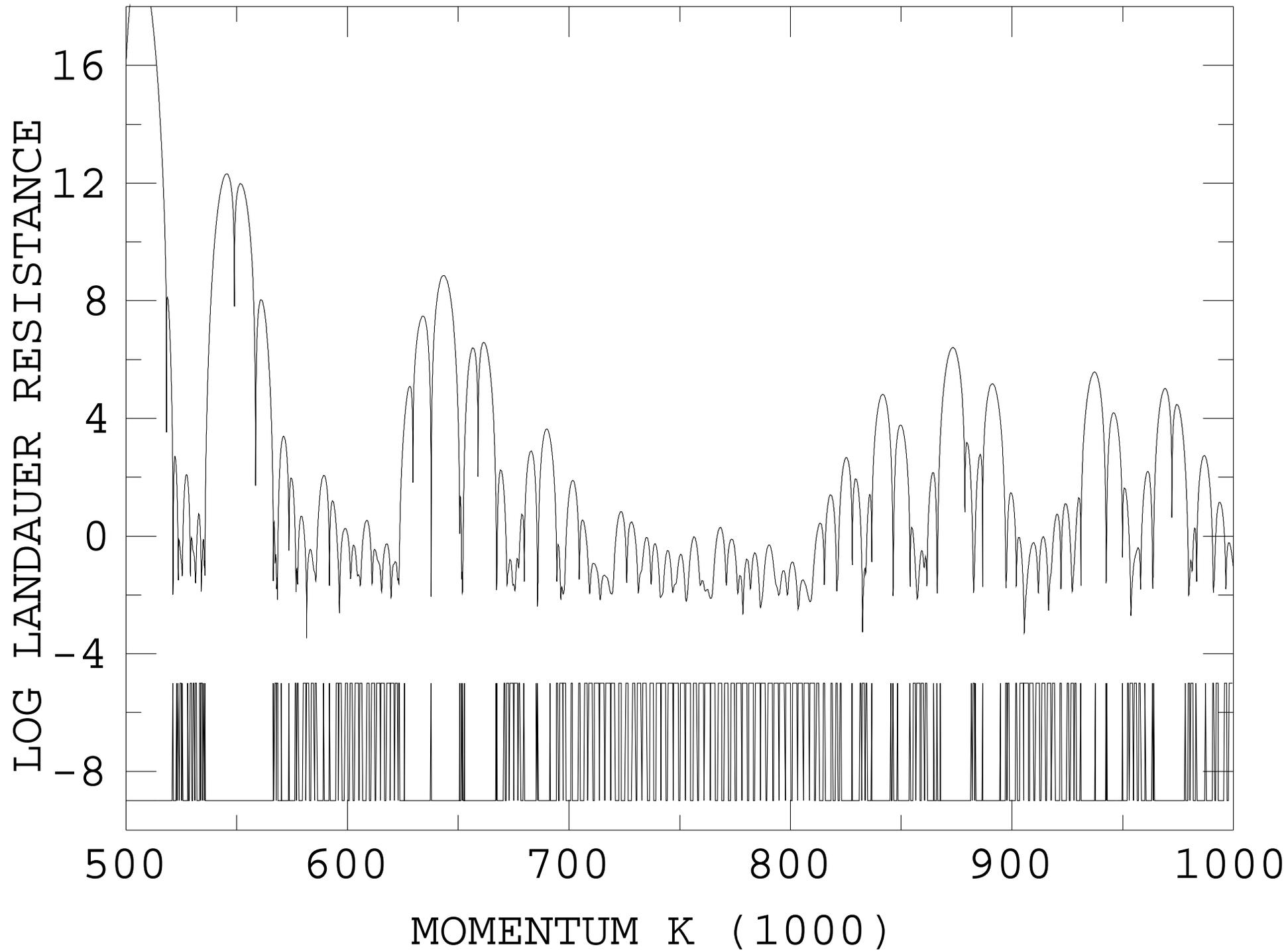
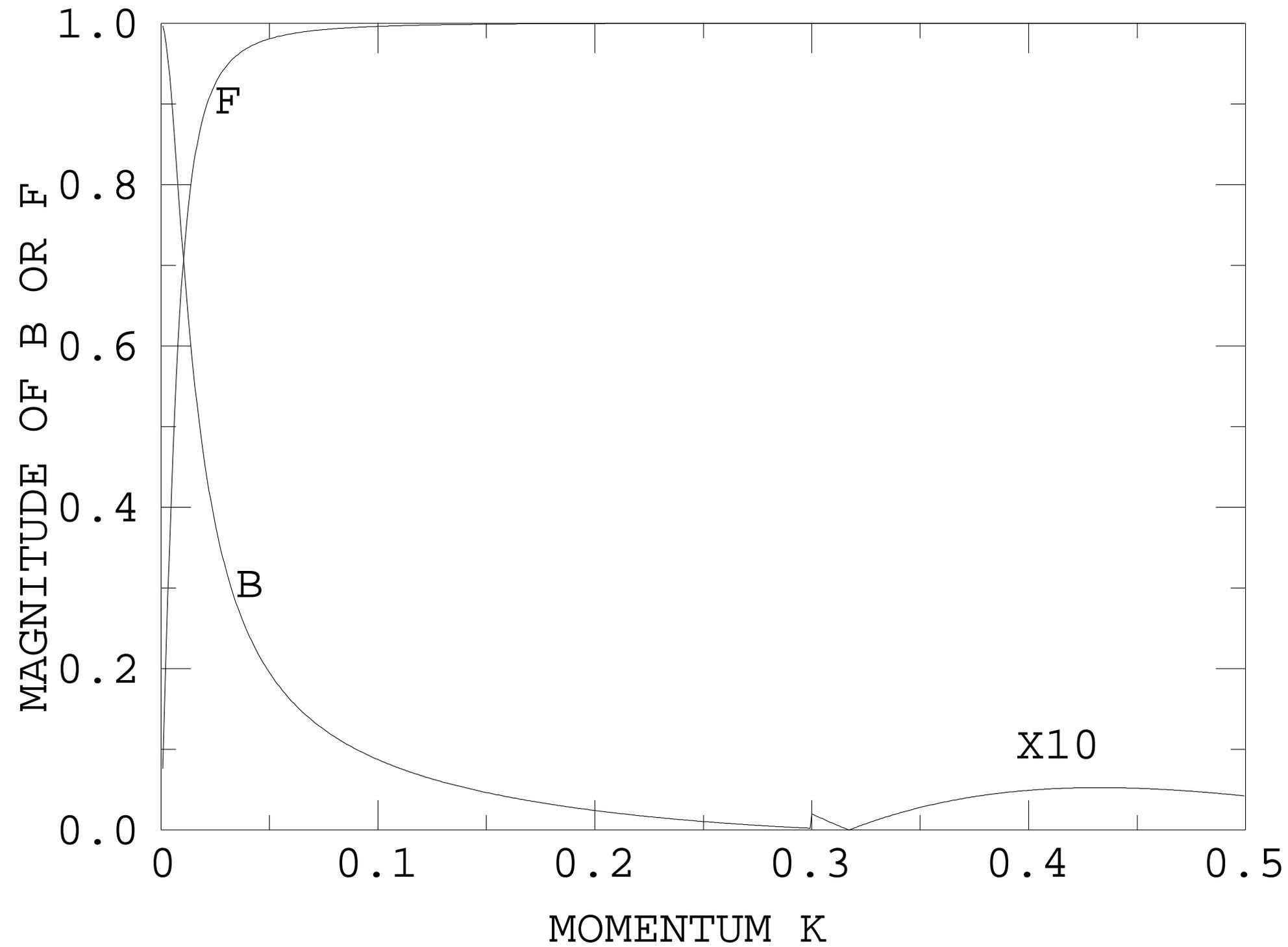
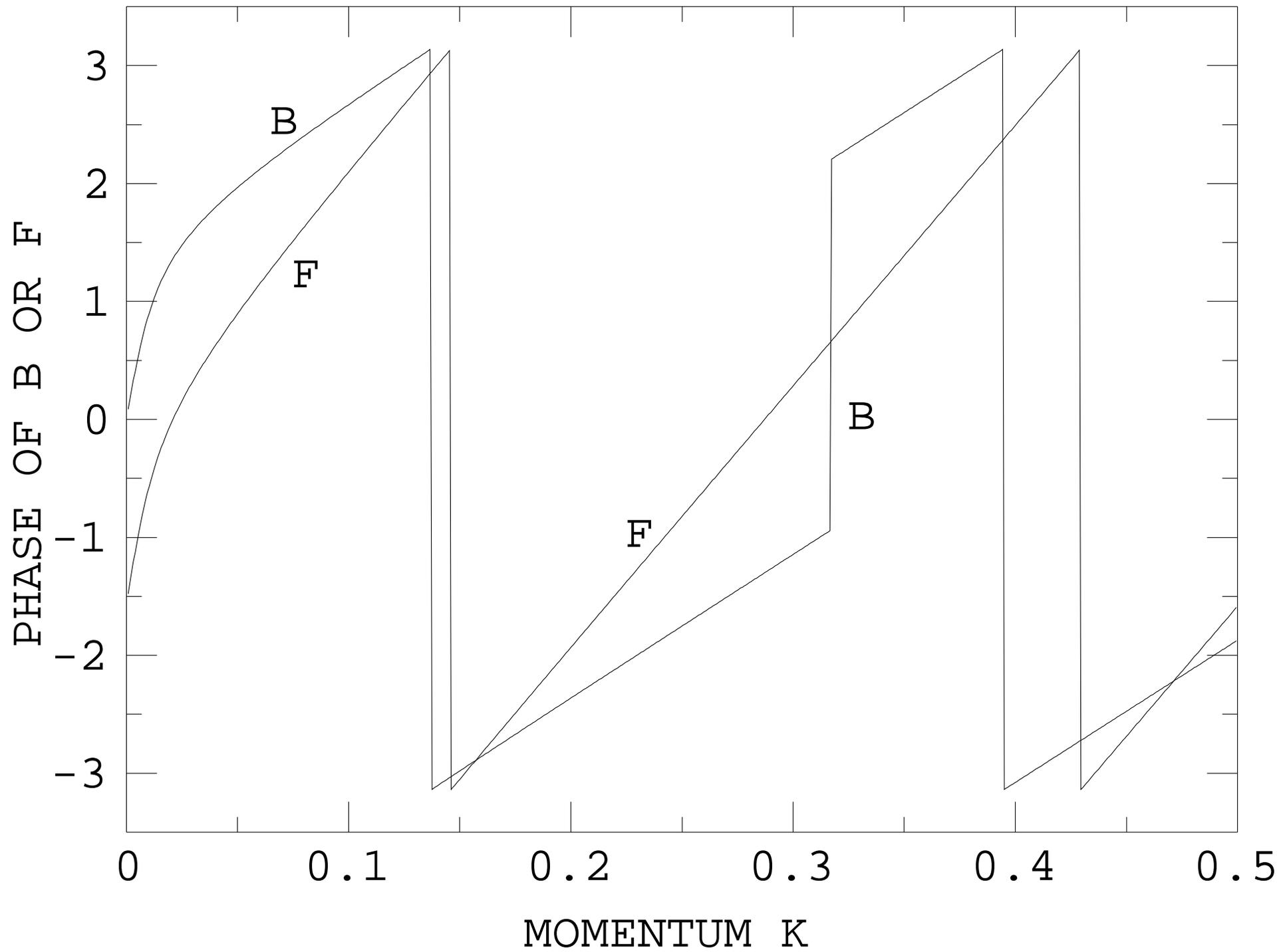
\end{document}